\pdfoutput=1
\documentclass[twocolumn,prb,showpacs,superscriptaddress,floatfix]{revtex4-1}
\usepackage[english]{babel}
\usepackage[ansinew]{inputenc}
\usepackage{times}
\usepackage{graphicx}
\usepackage{graphics}
\usepackage{amsmath}
\usepackage{amsfonts}
\usepackage{amssymb}
\usepackage{amsbsy}
\usepackage{epstopdf}
\usepackage{makeidx}
\usepackage{color}
\usepackage{pgf}
\usepackage{bm}
\usepackage{tikz} 
\usepackage[normalem]{ulem}
\usepackage{multirow,bigstrut}
\usepackage{tabularx} 

\newcommand{\be}{\begin{equation}}
\newcommand{\ee}{\end{equation}}
\newcommand{\bea}{\begin{eqnarray}}
\newcommand{\eea}{\end{eqnarray}}

\usepackage{multirow}

\usepackage{tikz,graphicx,dcolumn}
\usepackage{amssymb}
\usepackage{color}
\usetikzlibrary{shapes}

\DeclareRobustCommand\markerone{{\tikz{\node[draw,scale=0.5,circle,fill=black,black!20!black](){};}}}
\DeclareRobustCommand\markertwo{{\tikz{\node[draw,scale=0.5,circle,fill=red,red!20!red](){};}}}
\DeclareRobustCommand\markerthree{{\tikz{\node[draw,scale=0.5,circle,fill=blue,blue!20!blue](){};}}}

\usepackage{comment}
\includecomment{slow}

\makeindex
\begin{document}

\title{Machine Learned Interatomic Potential for Dispersion Strengthened Plasma Facing Components}

\author{E. L. Sikorski}
\affiliation{Center for Computing Research, Sandia National Laboratories, Albuquerque, New Mexico 87185, USA}
\author{M. A. Cusentino}
\affiliation{Material, Physical, and Chemical Science Center, Sandia National Laboratories, Albuquerque, New Mexico 87185, USA}
\author{M. J. McCarthy}
\affiliation{Center for Computing Research, Sandia National Laboratories, Albuquerque, New Mexico 87185, USA}
\author{J. Tranchida}
\affiliation{CEA, DES/IRESNE/DEC 13018 Saint Paul Lès Durance, France}

\author{M. A. Wood}
\affiliation{Center for Computing Research, Sandia National Laboratories, Albuquerque, New Mexico 87185, USA}
\author{A. P. Thompson}
\affiliation{Center for Computing Research, Sandia National Laboratories, Albuquerque, New Mexico 87185, USA}

\date{\today}

\begin{abstract}
Tungsten (W) is a material of choice for the divertor material due to its high melting temperature, thermal conductivity, and sputtering threshold.  
However, W has a very high brittle-to-ductile transition temperature and at fusion reactor temperatures ($\geq$1000~K) may undergo recrystallization and grain growth. 
Dispersion-strengthening W with zirconium carbide (ZrC) can improve ductility and limit grain growth, but much of the effects of the dispersoids on microstructural evolution and thermomechanical properties at high temperature are still unknown. 
We present a machine learned Spectral Neighbor Analysis Potential (SNAP) for W-ZrC that can now be used to study these materials. 
In order to construct a potential suitable for large-scale atomistic simulations at fusion reactor temperatures, it is necessary to train on \textit{ab initio} data generated for a diverse set of structures, chemical environments, and temperatures.
Further accuracy and stability tests of the potential were achieved using objective functions for both material properties and high temperature stability. 
Validation of lattice parameters, surface energies, bulk moduli, and thermal expansion is confirmed on the optimized potential. 
Tensile tests of W/ZrC bicrystals show that while the W(110)-ZrC(111) C-terminated bicrystal has the highest ultimate tensile strength (UTS) at room temperature, observed strength decreases with increasing temperature. 
At 2500~K, the terminating C layer diffuses into the W, resulting in a weaker W-Zr interface. 
Meanwhile, the W(110)-ZrC(111) Zr-terminated bicrystal has the highest UTS at 2500~K. 
\end{abstract}

\pacs{}

\maketitle

\section{\label{intro}Introduction}

\par 
High-purity, nano-crystalline Tungsten is a leading candidate for the divertor material in future fusion reactors and ITER due to its high melting temperature, thermal conductivity, and sputtering threshold\cite{federici2001,pitts2013}. However, W has a high brittle-to-ductile transition temperature ($\leq$ 473~K)\cite{Xie2015} and has been observed to undergo recrystallization and grain growth at ITER temperatures ($\geq$ 1000~K) \cite{Lang2021}. Taking into account off-normal events like edge-localized modes (ELMs), it is predicted the divertor surface will need to withstand up to 2573~K in realistic operating conditions\cite{Linke2019}. While the brittle character of W has often limited its application to wire and thin film geometries, experimental studies have found that adding carbide dispersoids (TiC, TaC, and ZrC) to W can lower the ductile-to-brittle transition temperature to that of a thin film\cite{Xie2015}.  
Additionally, it is hypothesized that the dispersoids prevent impurity (O, N, etc.) trapping at the grain boundaries, which leads to brittle fracture, by re-distributing the impurity concentrations\cite{lassner1999,wurster2013,kurishita2014}.  The dispersoids can also help to pin the grain boundaries and prevent recrystallizing at higher temperatures\cite{Xie2015,lang2018}. While dispersion strengthened W is certainly promising as a plasma facing material, the strengthening mechanisms and effects on thermomechanical properties are not fully understood.  Another concern is the effect of these dispersoids on hydrogen trapping, which may act as a sink for tritium implanting into the divertor. \par 
Given the complexity of experimentally studying plasma facing materials, simulation efforts are sought after to predict these mechanisms and outcomes.
Atomic scale modeling can be used to address some of the underlying questions surrounding the use of dispersoid strengthened tungsten as a plasma facing material, particularly at the Density Functional Theory (DFT) scale. Since first-principles methods can be used without experimental parameterization, they can offer predictive capabilities for material properties related to these dispersoid strengthening and impurity trapping mechanisms. 
Some DFT work has been done to look at interface stability for TiC\cite{ding2013,dang2015} and ZrC\cite{Zhang2020,Mukhopadhyay2022} dispersoid strengthened tungsten. A subset of these studies \cite{dang2015,Zhang2020} also looked at hydrogen and helium trapping at the tungsten-dispersoid interface, which was found to be a strong trap for both plasma species.  While this work can provide some insight on the properties of the tungsten-dispersoid interface, DFT can only resolve systems on the order of hundreds of atoms before encountering significant computational cost. With dispersoids on the order of tens of nanometers\cite{Xie2015}, fully resolving dispersion strengthened W, especially at the higher operating temperatures seen in a fusion reactor, is well beyond the reach of standard DFT. \par 

As a means to tackle these higher length/time scale phenomena, machine-learned interatomic potentials (MLIAPs) have demonstrated the ability to use DFT data to drive classical Molecular Dynamics (MD), which extends DFT accuracy to more experimentally relevant length scales and temperatures. With this capability, first-principles accurate predictions can be achieved for microstructures comprising billions of atoms\cite{nguyen2021billion,lu202186}. In this paper, we will use the Spectral Neighbor Analysis Potential (SNAP) methodology\cite{Thompson2015} to develop a quantum-informed W-ZrC interatomic potential. We chose ZrC as a model dispersoid due to its smaller grain sizes reported experimentally\cite{Xie2015} ($\approx$10-100 nm) as compared to TiC and TaC\cite{lang2018} ($\approx \mu$m), making it suitable for investigation with MD. The use of a well trained SNAP ML-IAP will allow an investigation of the effects of W-ZrC interfaces on mechanical properties at a range of temperatures inaccessible by DFT.
\par 
Developing the SNAP potential for use in production\cite{Braun2019} MD simulations is an iterative process, improving upon earlier, candidate potentials until one yielding the desired accuracy and performance is reached. SNAP model development consists of three primary phases: (1) constructing the training set, (2) optimizing the free variables, and (3) testing the performance of candidate potentials. In the MLIAP field, there are not yet established best practices in how to construct a training set. Towards greater transparency in MLIAP development, we include a detailed discussion of our training configurations included in this work. 
In practice, a potential is iteratively improved by identifying inaccuracies, supplementing the training set\cite{podryabinkin2017active,vandermause2020fly,farache2022active,Mishin2021}, and if needed applying constraints to ensure specific properties are captured\cite{Thompson2015,Wood2018,Wood2019,cusentino2020explicit}. 
For instance, when candidate potentials yield poor behavior in molecular dynamics simulations (phase 3), we can add structures to the training set (phase 1) and/or modify the optimization scheme (phase 2) to yield more physical behavior. While each phase will generally be discussed sequentially, we note that true SNAP development comprises multiple loops over the three phases.
While approaches to MLIAP development vary, phase 3 in particular is not widely adopted\cite{Deringer2021,Fu2022}. Instead, development often relies on accuracy with respect to the regression process during the potential optimization, even where active learning is employed. \cite{vandermause2020fly,smith2018less}
We aim to detail all three phases of the WZrC SNAP development herein as a paradigm for future MLIAP development in the field. 
\par
We will first summarize the SNAP formulation. Second, we will describe the configurations included in the training set. Third, we will discuss best practices for optimizing the variables (hyper parameters and group weights). Finally, we will examine the properties of the production potential and use it to investigate the thermomechanical properties of different W-ZrC interfaces.
\par


\section{\label{model} Potential Energy Model}
Over the past decade, numerous different MLIAP model forms have been proposed but can be generally classified by the use (or lack thereof) of a symmetry preserving descriptors as model input.
As one might expect, practitioners of MLIAP aim to preserve the intrinsic physical laws that govern observables such as energy, force, and stress. 
Most models satisfy rotation, translation, and permutation (of atomic indices) invariance. Different classes of MLIAP can be characterizes by whether these invariances are learned from the data or enforced explicitly in the mathematical structure of the descriptors.
Examples of the former include graph representations\cite{chen2019graph, zhang2018deep}, equivariant\cite{batzner20223, rackers2022cracking}, and message passing neural networks\cite{smith2019approaching, st2019message, zubatiuk2021development}.
Where these invariant properties are enforced in the descriptors, these are often referred to as descriptor-based MLIAP.  A summary and comparison of many of these has been recently published\cite{Zuo2020}.
The bispectrum components used in SNAP is one such descriptor, and has been demonstrated for use in single and multi-element forms\cite{cusentino2020explicit, Thompson2015,Wood2018}.
\subsection{\label{math-section}Spectral Neighborhood Analysis Potential}

\newcommand{\br}{{\bf r}}
\newcommand{\bu}{{\bf u}}
\newcommand{\bU}{{\bf U}}
\newcommand{\bB}{{\bf B}}
\newcommand{\balpha}{{\boldsymbol\alpha}}
\newcommand{\bbeta}{{\boldsymbol\beta}}

The total potential energy of a configuration of atoms can be written as a sum of atomic energies combined with an additional reference potential,
\begin{equation}
E({\bf r}^{N})=E_{ref}({\bf r}^{N})+\sum_{i=1}^{N}E_i({\bf r}^{N}),
\label{snapE}
\end{equation}
where $E$ is the total potential, ${\bf r}^{N}$ are the positions of the $N$ atoms in the configuration, $E_{ref}$ is the reference energy, and $E_i$ is the atomic energy of atom $i$.  This partitioning of total energy into atomic energies is not rigorous, but has proven to be effective in many MLIAPs, beginning with the pioneering work of Behler and Parrinello.\cite{Behler2007} 
The atomic energy of atom $i$ is expressed as a sum of the bispectrum components ${\bf B}_i$ for that atom weighted by regression coefficients 
\begin{eqnarray}
E_i({\bf r}^{N}) & = & 
\beta_{0{\nu_i}} + {\bbeta}_{\nu_i} \cdot ({\bB}_i -  {\bB}_{0\nu_i}) \,,
\label{eqn:e}
\end{eqnarray}
where $\beta_{0{\nu}}$ and the elements of the vector ${\bbeta}_{\nu}$ are constant linear coefficients for atoms of element ${\nu}$ whose values are determined in training. 
The vector ${\bB}_i$ is a flattened list of bispectrum components
for atom $i$, while ${\bB}_{0\nu}$ is the list of bispectrum components
for an isolated atom of type $\nu$.  By construction, the atomic
energy of an isolated atom is then $\beta_{0{\nu}}$.
The bispectrum components are real, rotationally invariant triple-products of four-dimensional hyperspherical harmonics $\bU_j$ \cite{Bartok2010}
    \begin{eqnarray}
        \label{eqn:b}
        B_{j_1j_2j}  &=& \bU_{j_1} \otimes_{j_1j_2}^j \bU_{j_2} \colon \bU_j^* \,, \end{eqnarray}
where symbol $\otimes_{j_1j_2}^j$ indicates a Clebsch-Gordan product of two matrices of arbitrary rank, while $:$ corresponds to an element-wise scalar product of two matrices of equal rank. The total hyperspherical harmonics for a central atom $i$ are written as sums over neighbor contributions,    
    \begin{eqnarray}
        \label{eqn:u}
        \bU_j &=& \bu_j({\bf 0}) + \sum_{k \in \mathcal{N}(i)}~f_c(r_{ik})~g_c(r_{ik}) w_{\nu_k} \bu_j(\br_{ik}) \,,
    \end{eqnarray}
where the summation is over all neighbor atoms $k$ within a cutoff distance $R_{\nu_i \nu_k}$ of atom $i$.  The hyperspherical harmonics $\bu_j$ are described below.  Atoms of different chemical elements are distinguished by the element weights $w_{\nu}$.  The radial cutoff function $f_c(r)$ ensures that atomic contributions go smoothly to zero as $r$ approaches $R$ from below. Similarly the inner cutoff function $g_c(r)$ ensures that
contributions go smoothly to zero at very short separations, which prevents unphysical attractive energies for high density clusters
that are not represented in training.  Both of these
radial cutoff functions are composed from cosine half-periods,
\begin{eqnarray}
      f_c(r)   & = & h(-\frac{r}{R}) \nonumber \\
  g_c(r) & = & h(\frac{r-(S+D)}{2D}) \nonumber \\
    h(z) & = & 1, \mspace{154mu}  z < 0 \nonumber \\
   & = & \frac{1}{2}(\cos{(\pi z)} + 1), \,\,\, 0 \leq z \leq 1 \\
        & = & 0, \mspace{154mu} z > 1 \nonumber,
\label{eqn:inner}
\end{eqnarray}
where $R$, $S$, and $D$ are hyper-parameters that are 
optimized separately for each pair of chemical elements. The ZBL\cite{zbl} repulsive potential was overlayed with SNAP with a switching function range from 0.9 to 1.4 $\mbox{\AA}$ to govern the physics at short interatomic distances.

The hyperspherical 
harmonics $\bu_j$ are also known as Wigner matrices, each of rank $2j+1$, and the index $j$ can take half-integer values $\{0, \frac{1}{2},1,\frac{3}{2},\ldots\}$.
They form a complete orthogonal basis for functions defined on $S_3$, the unit sphere in four dimensions.\cite{Varshalovich1988} The relative position of each neighbor atom ${\br}_{ik}= (x,y,z)$ is mapped to a point on $S_3$ defined by the three polar angles $\psi$, $\theta$, and $\varphi$ according to the transformation $\psi = \pi r/r_0$, $\cos\theta = z/r$, and $\tan\varphi = x/y$. The bispectrum components defined in this way have been shown to form a particular subset of third rank invariants arising from the atomic cluster expansion \cite{Lysogorskiy2021}. The vector of descriptors ${\bB}_i$ for atom $i$ introduced in Eq.~\ref{eqn:e} is a flattened list of elements $B_{j_1j_2j}$ restricted to $0 \le j_2 \le j_1 \le j \le J$, so that the number of unique bispectrum components scales as $\mathcal{O}(J^3)$.

The choice of both inner and outer radial cutoff parameters and element weights are taken as hyper-parameters of the MLIAP training as they directly affect the calculation of these descriptors.
This will be discussed further after addressing the other phases of the overall training routines developed here.

\subsection{\label{training-set-construction} Constructing the Training Set}
\begin{slow}

MLIAP development generally relies on an understanding of the material system, application or domain expertise (DE) to construct the training set. Emerging strategies also show promising results towards automating training set configuration, for example with active learning \cite{vandermause2020fly, farache2022active, podryabinkin2017active}. These strategies to move beyond DE (BDE) help expand the expected application space where the MLIAP can be used accurately, since MLIAPs often exhibit poor behavior when extrapolating to new atomic configurations\cite{Mishin2021,Unruh2022,MontesdeOcaZapiain2022}. While MLIAPs have shown reasonable extrapolation behavior in some cases\cite{Rosenbrock2021,Lysogorskiy2021}, MD configurations very dissimilar to the training set will nevertheless have lower accuracy than those similar to the training set\cite{Mishin2021}. With a desire to simulate plasma-material interactions where temperatures can reach up to 2500~K, the MLIAP developed in this work may quickly leave configuration space conceived by expert users during MD. In order to mitigate these errors, we have used a variety of strategies to construct a BDE W-ZrC training data set. This section will provide an overview of the approximately $10^{4}$ configurations in the training set by describing the overarching types of structures and how they contribute to the MLIAP performance. We provide detailed discussion about why different types of structures were included to improve transparency in MLIAP development and reporting. The discussion begins with DE structures and moves progressively towards BDE structures.
\par 

The DFT calculations for all sources of configurations were performed using the Vienna Ab initio Simulation Package (VASP)\cite{Kresse1996}. Spin-polarized generalized gradient (GGA) exchange-correlation functional was used with the Perdew$\mbox{-}$Burke$\mbox{-}$Ernzerhof (PBE) formulation\cite{Perdew1996}. Plane-wave basis sets were implemented utilizing projector-augmented wave (PAW) pseudo-potentials. Bulk, ground state and USPEX\cite{Oganov2006, Oganov2011, Lyakhov2013} structures were simulated with a cutoff energy of 540 eV and 8$\times$8$\times$8 k-points. Ground state surfaces and interfaces were optimized with a minimum of 420 eV and 4$\times$4$\times$1 k-points, with k-point sampling adjustments respective to the simulation cell geometries. Ab initio Molecular Dynamics (AIMD) simulations (represented by the stars in Figure \ref{fig:training_set}) were run with a minimum of 540 eV cutoff energy and sampled at the gamma point for computationally reasonable throughput. The structures were visualized using Visualization for Electronic and STructural Analysis (VESTA)\cite{Momma:db5098}.

\begin{figure}[htb]
    \centering
    \includegraphics[width=0.45\textwidth]{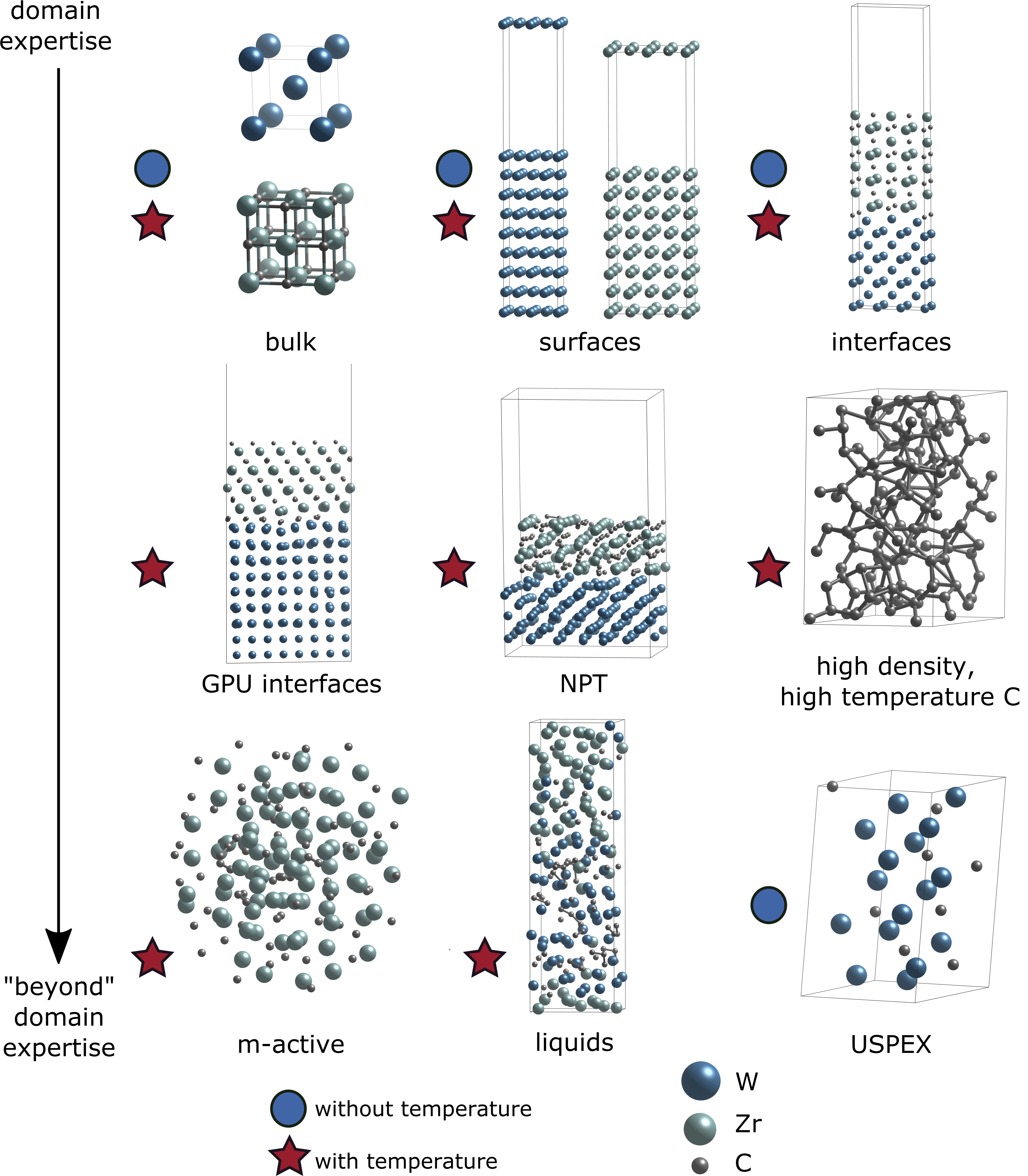}
    \caption{Overview of structures included in the W-ZrC training set. Circles and stars are used to indicated structures simulated without and with temperature, respectively. The top row depicts more traditional DFT structures that a domain expert would include. Moving towards the bottom of the figure, the structures begin to move away from domain expertise which include disordered and high formation energy structures.}
    \label{fig:training_set}
\end{figure}

A schematic overview of the structures in the training set is shown in Figure \ref{fig:training_set}. The DE structures begin with unit cells of W and ZrC. From a traditional DFT perspective, unit cells describe the behavior of the bulk, or regions sufficiently far from interfaces or surfaces, due to periodic boundary conditions (PBCs). To study surfaces, grain boundaries, voids, or other regions where free space may be present, we included surfaces with low Miller indices for both W and ZrC. These surface structures also help the potential fit to stable terminations of W and ZrC. The surface structures in this training set used the asymmetric slab model, where layers on one side of the slab are frozen (representing the bulk) and the others are allowed to relax. To optimize the potential for modeling ZrC dispersoids in W, we need to include expected interfaces between W and ZrC. The training set includes combinations of W (100), (110), and (111) with ZrC (100), (110), and (111).  Orientations were chosen based on prior DFT calculations\cite{Zhang2020,Mukhopadhyay2022} of interface stability.
\par 
These DE structures provide the potential with information on the most stable, ground state configurations. However, they provide little information on how the potential should capture atomic displacements due to high temperatures nor any other possible reconstructions. For this reason, snapshots from AIMD simulations were included for the bulk, selected surfaces, and selected interfaces at room temperature (300~K) and high temperature (1000~K - 5000~K). 


\begin{figure*}[htb]
    \centering
    \includegraphics[width=1.0\textwidth]{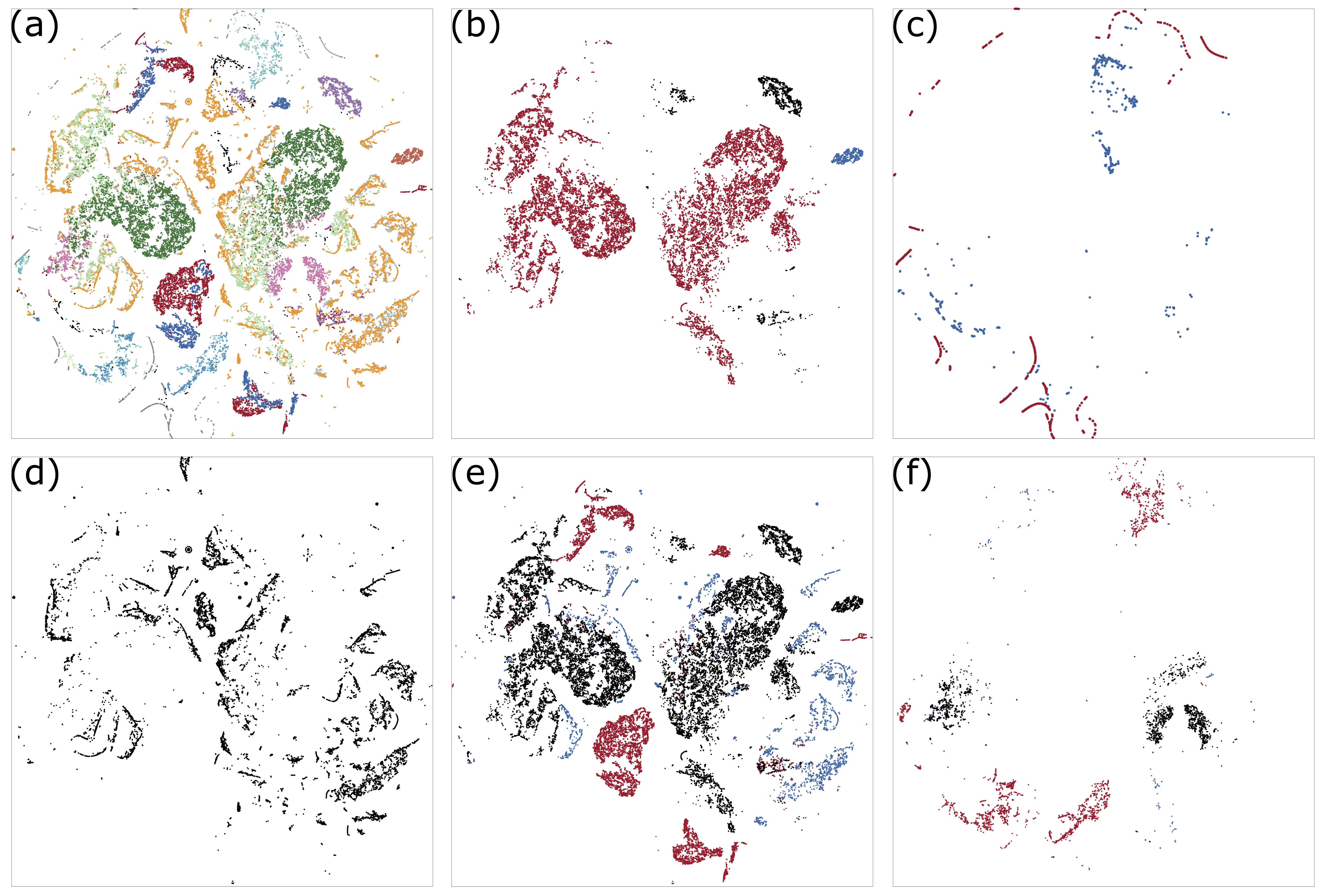}
    \caption{
    Visualization of the training set coverage on the bispectrum descriptor space in 2D using distance preserving t-SNE analysis. 
    (a) All data, labels in Figure \ref{fig:tsne-all-labels}
    (b) Labeling by constituents : \markerone{} W, \markertwo{} ZrC, \markerthree{} C 
    (c) Labeling for ground states :  \markertwo{} EOS, \markerthree{} unit cells defects, etc.
    (d) Labels for AIMD : \markerone{} $T=300$~K
    (e) Labels for High Temperature AIMD : \markerone{} W/ZrC/C, \markertwo{} $>200$atoms, \markerthree{} interfaces
    (f) Labels for beyond domain expertise : \markerone{} Liquids, \markertwo{} USPEX generated variable composition, \markerthree{} M-active generated.
    }
    \label{fig:t-sne}
\end{figure*}


\par 
In the second row of Figure \ref{fig:training_set}, we move to structures that can still be set up by a domain expert, but utilize some additional strategies to create more unique structures. While we can make informed decisions on what interfaces to include, it is nearly impossible to include all possible interface reconstructions, especially at high temperature. Additionally, standard, computationally feasible DFT cell sizes limit the amount of reconstruction that can occur at interfaces. These finite size effects only allow for low energy reconstructions since any structural changes are forced to occur regularly throughout space (due to periodic boundaries). At the same time, we want to pay special attention to certain interface reconstructions, as Mukhopadhyay et al. \cite{Mukhopadhyay2022} showed they play a significant role in the stability of a W-ZrC interface.
\par
If we can reach larger system sizes and thus larger interfacial area, we can train  on higher energy reconstructions and/or those that may happen less frequently.  We leveraged VASP GPU capabilities to run AIMD simulations of interfaces with large atom counts (up to $\approx$500 atoms).
These interfaces were run at 1000~K, 3000~K, and 5000~K. While 5000~K is above the melting temperature of both W and ZrC, this provides the potential with information on very high energy reconstructions and intermixing of all chemical species. 
It is also important to include interface training data for Miller index combinations that we want to run the final potential on. Without such training, running a structurally or chemically extrapolative interface can lead to an amorphous structure at high temperature which was observed for early candidate SNAP potentials without such training data.
\par 
MLIAP development also benefits from structures opposing traditional DFT approaches. For instance, DFT studies often strive to minimize strain at the interface in order to find ground state structures. This is usually done by modifying the constraints on the system, e.g. by keeping the volume constant whilst allowing the dimensions to optimize. In contrast, an MD potential needs to be capable of exploring not only ground state structures but any likely behavior at the desired temperature, especially for the high temperature application of this potential. We also note that since MLIAPs are trained not only on energy but also the corresponding forces, thus any structure can provide useful training information. To provide the potential with some information about how W and ZrC may try to accommodate each other at an interface and at higher energy, we included isobaric AIMD snapshots of a 19$\times$19 $\mbox{\AA}$ supercell at 1000~K. This gives the potential insight into how variations in the surface area at an interface affect both the energy and the forces.
\par 


After testing of SNAP candidates, it was evident that additional training was needed to describe C-C interactions, especially at small interatomic distances. As the atomic diffusion will be dominated by the lightest atoms\cite{Feenstra1999}, we can anticipate our lightest element, C, may leave the configuration space included in the training set most easily\cite{Deringer2017}. Such motion can, for example, launch neighboring C atoms to short interatomic distances that should lie on the repulsive energy wall. To combat this behavior, additional structures were included to provide examples of physical C-C behavior at short distances. We note that at very short distances DFT can also fail. Additional strategies towards physical short distance C-C behavior are discussed in Section \ref{optimize}.
\par 
To form this additional C training, an initially graphite unit cell was used to create a C supercell (128 atoms). The C supercell was compressed to 70\% of its original volume and run via AIMD at 5000~K. Every fifth snapshot from a 500 fs (1 fs timestep) run was included in the training set. With this data, the potential is trained on possible C-C configurations at short distances and the corresponding energies and forces as compared to ground state structures. This helps us to maintain 1 fs timesteps during potential use without C atoms migrating/exploring away from training space.

\par
In the third row of Figure \ref{fig:training_set}, we begin with manual active (m-active) learning structures. These m-active structures were identified during MD testing of SNAP candidates when the developers were unsure of the accuracy of a certain behavior. Namely, the W-ZrC potential has a tendency to form C trimers. To ensure such configurations were included in the training set so that the behavior could be corrected if necessary, DFT-sized cells including trimers were carved out of the MD simulations containing poorly behaved dynamics. Additionally, a small section of a ZrC dispersoid and adjacent W was carved out to account for a geometry beyond a planar interface.
\par 

Liquid structures provide another method to include configurations of W, Zr, and C that may not be initially considered by a DFT domain expert. Though these configurations have been gaining prevalence in MLIAP training sets\cite{Thompson2015,Wood2018,Wood2019,Mishin2021,Nikoulis2021,MontesdeOcaZapiain2022}, their contribution to solid-state applications is not necessarily intuitive from a DFT perspective. To expedite liquid generation, DFT-sized interfaces were run in MD using a SNAP candidate at high temperature. An output MD snapshot was then returned to AIMD for simulation at 1000~K, 3000~K, and 5000~K.  


\par 

The final type of structures in the training set were generated using USPEX, a program that searches for lowest energy structures using a genetic algorithm (GA)\cite{glass2006uspex}. USPEX focuses on optimizing the GA (e.g. initializing the structure population and producing children structures) while allowing the user to choose from several atomistic codes to obtain the energy of each structure. With this flexibility, we can search for structures using energies obtained from either DFT or MD.
\par 
For the structures produced by coupling USPEX with DFT, we utilized the variable composition feature. This allows the user to pick elements, e.g. W and C, and USPEX will produce a range of stoichiometries and different configurations for each stoichiometry. We included composition sweeps for each possible binary combination of W, Zr, and C. Instead of following the genetic algorithm progression, we took up to 200 structures for each binary composition that USPEX generated as an initial population. An additional 39 USPEX structures were included for C to provide more descriptions of C-C interactions. For throughput, these variable composition sweeps used lower DFT accuracy (e.g. 350 eV cutoff energy). After the structures were generated, they were optimized again using DFT at the same level of accuracy as the other bulk, ground state training data (540 eV and 8x8x8 k-points). For each variable composition structure, the first 50 ionic steps (or up until the point of electronic step failure) were included in the training set. These composition sweeps are particularly helpful when the developer may not know what types of compositions may form \emph{a priori}. In the case of W-ZrC, while tungsten carbide may form in the vicinity of the W-ZrC interface\cite{Xie2015,lang2018}, limited information is known about the structure, composition, and atomistic formation mechanism. Without much guidance on what tungsten carbide structures to include, the USPEX WC composition sweep is able to cover a DE gap\cite{xie2016effects}. 
\par 
By coupling MD with USPEX, we can include an additional flavor of active learning. For these structures, we utilized SNAP candidates with low errors and promising properties to search the potential energy surface. This provided fast, invaluable information on what areas of configuration space a SNAP candidate would attempt to enter, given the required energy to overcome any saddle points. After finding structures the SNAP candidate considered to be stable, the structures were returned to DFT for energy and force evaluation and added to the training set. Active USPEX structures were included for stoichiometric ZrC, compositions including W and C, and compositions including W, Zr, and C.

While we can attempt to form an intuition on how different types of training data may have unique or redundant contributions to an MLIAP, this does not provide a metric on how to build from scratch or improve upon an existing training set. To obtain a qualitative metric, we can visualize how our training configurations appear in SNAP descriptor space using t-distributed stochastic neighbor embedding (t-SNE)\cite{VanderMaaten2008}, as shown in Figure \ref{fig:t-sne}. This method works to preserve high dimensional structures while allowing for 2D visualization, so in our case the hypersphere coordinates can be visualized on a circle. This helps us understand how different training set configurations contribute to descriptor coverage and the similarity between groups.
\par 
Figure \ref{fig:t-sne}(a) shows the t-SNE of the W-ZrC training set broken down by configuration types. Due to the sheer number of configurations, only $\approx$10\% of most configuration types are visualized to retain interpretable shapes for each configuration type. Since each point represents a local atomic environment, smaller unit cells provide less points. For this reason, 100 \% of the ground state and USPEX composition sweep configurations are visualized. We can first break up this t-SNE plot to show the regions distinguished by atom types i.e. W, ZrC, and C AIMD configurations (Figure \ref{fig:t-sne}(b)). In Figure \ref{fig:t-sne}(c), we can see that ground state DFT data sparsely covers the circle. Meanwhile, by including room temperature AIMD snapshots alone (Figure \ref{fig:t-sne}(d)), we can cover significantly more configuration space and begin to populate the circle. The high temperature AIMD configurations Figure \ref{fig:t-sne}(e) cover even more space, with especially unique contributions from configurations with $>$200 atoms that overlap minimally with other configuration types. Figure \ref{fig:t-sne}(f) shows the BDE type m-active, liquid, and USPEX configurations. We can see that these BDE configurations occupy different regions of descriptor space than the pure W, ZrC, or C configurations. The USPEX variable composition structures in particular can reach regions of the descriptor space not covered by DE structures. Considering the up to 2500~K application space, training on ground state and room temperature AIMD alone would leave a good amount of descriptor space unexplored. With this t-SNE, we can visually see how our application space for our MLIAP has been expanded by adding unique high temperature AIMD and BDE configurations.

\par
\par
\par


\par 
\par
\par

\end{slow}

\subsection{\label{optimize}Optimization Methodology}
\begin{slow}

Once the training set has been determined to be sufficient, additional properties within the SNAP parameterization need to be optimized. SNAP considers two types of variables or parameters when determining the bispectrum coefficients: hyper-parameters and group weights, explained in greater detail below. Optimizing these variables often yields a phase space of 20 dimensions or more. In order to reach optimal combinations of SNAP variables, we use the single objective genetic algorithm (GA) implemented in Dakota\cite{dakota} to search this space. The workflow for connecting Dakota, FitSNAP, the DFT training data, and LAMMPS is depicted in Figure \ref{fig:fitsnap}. 
LAMMPS is a general purpose simulation engine for particle dynamics\cite{Plimpton1995,thompson2022lammps}, and may of its efficient parallelized functions can be called from external codes, which we take advantage of herein. 
Dakota generates hyper-parameter and group weight values for each candidate that FitSNAP uses along with the DFT training data to generate bispectrum coefficients. FitSNAP calls LAMMPS to convert atom positions into bispectrum components for each configuration and then calculates the energy and force errors, i.e. how much the SNAP predictions differ from the energy and force values in the training set, using linear regression. FitSNAP returns these errors to Dakota. Material property and stability objective functions consisting of results from small LAMMPS calculations with the SNAP candidate are also returned to Dakota. Dakota uses the energy, force, and objective function errors to decide which candidates will participate in reproduction for the next GA generation. With the new generation of parameters, FitSNAP creates a new generation of SNAP candidates. This process repeats until the developer arrives at suitably optimized candidates for the respective Dakota run. This whole workflow, including modifying training data, is generally looped over many times before arriving at a production potential.

\subsubsection{\label{SNAP_variables}SNAP variables}

The SNAP hyper-parameters include the radial cutoffs $R$ (one for each element), the element weights, and inner cutoff parameters $S$ and $D$ from Eq. \ref{eqn:inner}. The SNAP group weights correspond to the groups that the training set is broken down into. The group weights will tell FitSNAP how much to prioritize the energy and force errors of each group. 
The group weight feature allows the user to fine-tune the importance of different training groups. For example, the user can  place increased importance on high-accuracy ground state configurations while  simultaneously lowering the weights on high energy structures without interfering with the potential energy surface construction. If all training data were to be weighted equally, this could cause the fit to skew towards reproducing the energies of outliers with the most fidelity. Similarly, without group separation, it would be difficult to include both traditional DFT and AIMD training data. Due to the differences in both the governing physics and the ionic steps desired between traditional DFT and AIMD simulations, one set of cutoff energies and k-points cannot always be applied to both DFT and AIMD and yield optimal results. Specifically, DFT calculations can generally converge with lower cutoff energies but require higher k-points. Meanwhile, AIMD simulations need higher cutoff energies to include higher energy occupied states, but are often sampled at the gamma point due to lower reciprocal space dimensions and to balance computational cost. Some MLIAPs have addressed this discrepancy by recalculating AIMD snapshots at the accuracy of the rest of the training set \cite{Zuo2020} to maintain a consistent potential energy surface. However, this adds great computational expense to informing the potential with AIMD. By splitting DFT and AIMD simulations into separate groups and including objective functions demanding high accuracy from properties of interest, Dakota\cite{dakota} and FitSNAP can collectively arrive at SNAP solutions that both maintain high fidelity to traditional DFT calculations and are informed by AIMD. 
\par 

\begin{figure}[htb]
    \centering
    \includegraphics[width=0.45\textwidth]{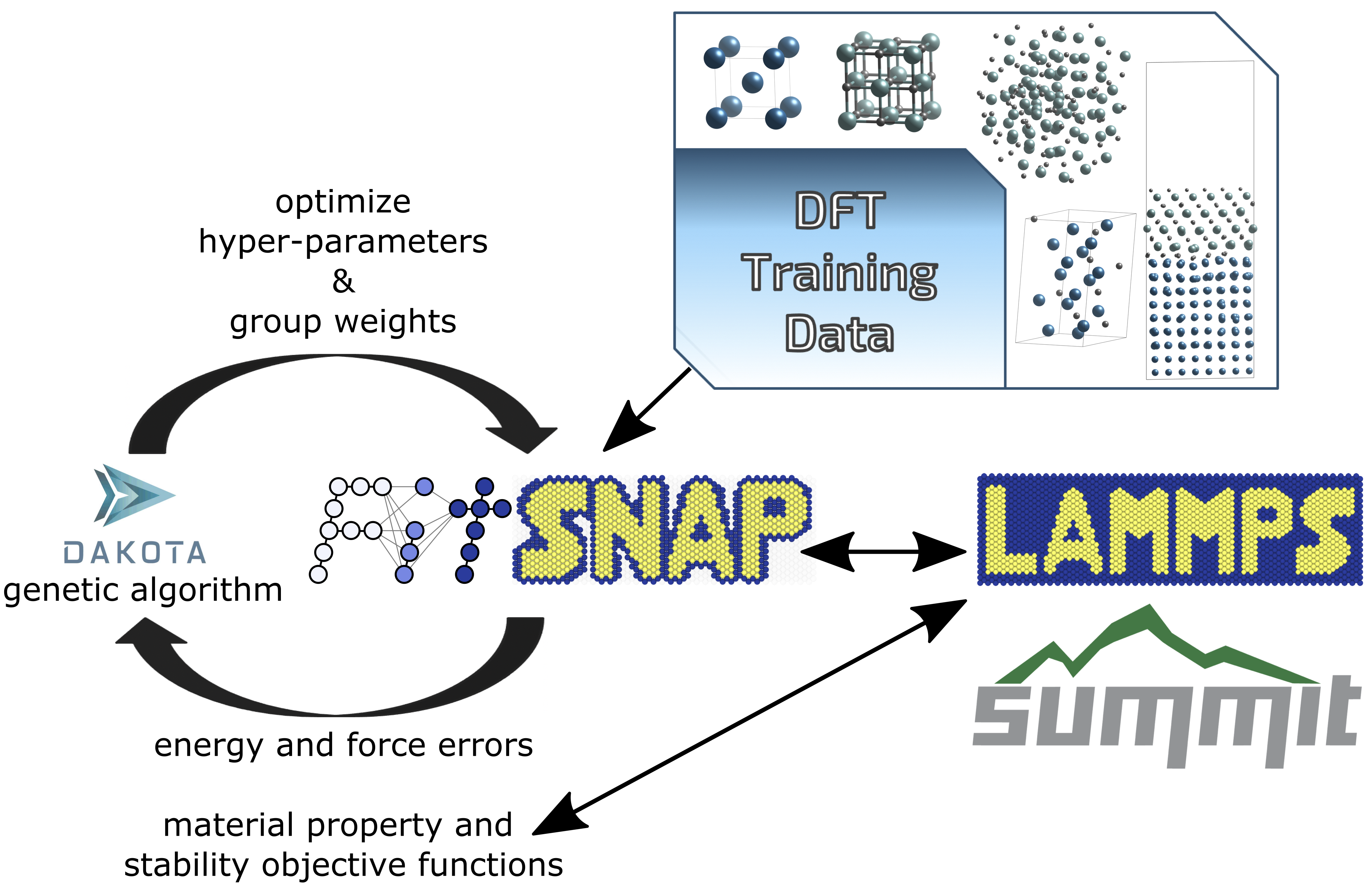}
    \caption{Schematic of the workflow for SNAP potential Development. First, FitSNAP reads in DFT training data. Using hyper-parameters and group weights assigned to the respective candidate by Dakota, FitSNAP determines the bispectrum coefficients. FitSNAP then calls LAMMPS to evaluate the forces and energies for the training configurations as predicted by the SNAP candidate. The difference between these SNAP predictions and the DFT values are fed to Dakota, along with errors from material property objective functions. Dakota uses the error information to determine which candidates should participate in reproduction and uses GA strategies to produce the next generation of candidates. Ultimately, the potential can be used to run large scale MD simulations on a machine such as Summit.}
    \label{fig:fitsnap}
\end{figure}

\par 
\subsubsection{\label{objective functions}Objective Functions}

Using objective functions helps ensure that our SNAP potential yields optimal material properties and stable, physical dynamics at high temperatures. The production SNAP potential was fit using eight material property objective functions and four stability functions.
Five objective functions were initially included to minimize the error for material properties: (1) W (100) and (110) surface energies, (2) ZrC (100) and (110) surface energies, (3) selected W-ZrC interface energies, (4) the W bulk modulus, and (5) the ZrC bulk modulus. The surface energies were calculated using 
\begin{equation}
    E_{surf} = \frac{E_{slab}-E_{bulk}}{2A} \quad ,
\end{equation}
where $E_{slab}$ is the total energy of slab, $E_{bulk}$ is the total energy of the bulk system with the same number of atoms as the slab, and $A$ is the surface area. The surface energy objective functions help ensure the formation energy for each surface is in the correct order and of reasonable magnitude, especially after geometry relaxation. This differs from the standard regression implemented in FitSNAP that only checks whether single-point (no geometry optimization) SNAP calculations match DFT for the training set structures. Similarly, interface energies were calculated using\cite{Zhang2020}
\begin{equation}
    E_{int} = \frac{E^{int}_{tot}-N_{W}E^{bulk}_{W}-N_{ZrC}E^{bulk}_{ZrC}}{S}-E^{surf}_{W}-E^{surf}_{ZrC} ,
\end{equation}
where $E^{int}_{tot}$ is the total energy of the interface, $N_{W}$ is the number of W atoms in the interface, $E^{bulk}_{W}$ is the bulk W energy per atom, $N_{ZrC}$ is the number of Zr-C pairs in the interface, $E^{bulk}_{ZrC}$ is the bulk ZrC energy per Zr-C pair, and $S$ is the area of the interface.
We note that while previous DFT work has used rigorous relaxation procedures to reach ground state interfaces \cite{Mukhopadhyay2022}, we performed simpler interface geometry optimizations. The purpose of the interface objective function is thus to check whether the SNAP candidate, following identical constraints during geometry optimization to the DFT reference, reaches an interface energy of the same magnitude.
\par
Fitting to energies, forces, and the aforementioned objective functions was insufficient to ensure thermal expansion of the correct magnitude or even the correct sign. Objective functions were added for both W and ZrC to check whether the thermal expansion from room temperature to 3000~K and 2793~K, respectively, matched experiment. The thermal expansion was calculated as\cite{Kostanovskiy2018}
\begin{equation}
    \bar{\alpha} = \frac{l_{T}-l_{0}}{l_{0}\big(T-293\big)} \quad ,
\end{equation}
where $l_{T}$ is the lattice parameter at the temperature of interest, $l_{0}$ is the lattice parameter at 293~K, and $T$ is the temperature of interest. 
\par 
Some SNAP candidates would incorrectly predict the most stable structure of W and ZrC, e.g. predicting face centered cubic (FCC) structure instead of body centered cubic (BCC) structure for W and predicting space group 198 instead of 225 (rocksalt) for ZrC. While including FCC W structures in the training set was sufficient to correct the SNAP candidate to form BCC W, many SNAP candidates formed space group 198 ZrC even after it was added to the training set. To better prevent this behavior, an objective function was added to return an error if the SNAP candidate predicted space group 198 as more stable than 225 for ZrC.
\par 
In addition to the material property objective functions, four stability objective functions were added to improve the performance of the candidates in MD simulations. These included (1) mitigation of atomic cluster formation $<$1.75 $\mbox{\AA}$ apart, (2) graphite bulk modulus, and radial distribution functions (RDFs) for (3) W and (4) ZrC surfaces at high temperature. As mentioned in Section \ref{training-set-construction}, it is both obscure and practically difficult to include a significant amount of training with short interatomic distances. For example, the majority of the C-C distances in the compressed C training data is at $\approx$ 1.4 $\mbox{\AA}$, with very limited training down to $\approx$ 0.9 $\mbox{\AA}$. Without much training below 1.4 $\mbox{\AA}$, SNAP(and any ML-IAP) is not fundamentally constrained to physical behavior at those distances. Also, turning off the SNAP contribution at short interatomic distances (Eq. \ref{eqn:inner}) and including the ZBL reference potential promotes but does not ensure desired physical behavior.
For example, during MD testing of early SNAP candidates, C atoms would attempt to move to positions $<$1.0 $\mbox{\AA}$  away from other C atoms. Subsequently, nearby W and Zr atoms would enter this region of multiple atoms sitting $<$1.0 $\mbox{\AA}$ apart. To mitigate this behavior, we added a stability objective function to return errors for cluster formation. This objective function begins with a small ($\approx$2000 atom) test of ZrC embedded in W. The test uses the SNAP candidate to minimize the small dispersoid geometry and run NVT at 2000~K. LAMMPS is used to sort any atoms less than 1.75 $\mbox{\AA}$ apart into clusters. The test then applies a penalty value based on the size of the largest cluster, so that as the Dakota run progresses it will move away from variables that result in this unphysical clustering. 
\par 
Similarly, a graphite bulk modulus objective function was added to improve the stability of C. This test has any candidate that produces a positive graphite bulk modulus value return a negligible error, while candidates that produce negative bulk modulus values are given an error proportional to the magnitude of that bulk modulus.
\par 
During the optimization process, while successively good candidates often had excellent bulk behavior, they had poorer performance for surfaces and interfaces, especially at high temperature. Furthermore, interfaces appeared to trigger clustering. For these reasons, objective functions were added to test whether the SNAP candidates could reproduce the AIMD RDFs for W and ZrC surfaces near or above their melting temperature. The ZrC RDF objective function was set to return an arbitrarily large error if it resulted in a cluster with more than 4 atoms. Otherwise, both the W and ZrC RDF objective functions returned the sum of the differences between the SNAP and AIMD RDF for each bin.

\par 
With an increasing number of objective functions, the computational time for each Dakota run quickly adds up. Namely, for each candidate, FitSNAP needs to determine the bispectrum coefficients for the given variables, and all of the objective functions need to be run. A tradeoff has to be made between providing each candidate one node so that more candidates can be evaluated at once and providing each candidate multiple nodes to speedup objective function evaluation but possibly reduce the total number of evaluated candidates. To expedite the candidate throughput, only eight objective functions (the original five material property objective functions, the W and ZrC surface RDFs, and the ZrC space group) were evaluated for every single candidate. The remaining thermal expansion and stability objective functions were set to run only if the sum of the previous force, energy, and objective function errors were below some threshold value. This allowed each node to continue on to the next candidate if the current candidate was already exhibiting poor energy, forces, and/or material property errors. 

\subsubsection{\label{GA}Genetic Algorithm Optimization}

While separating the training into different groups allows us to include more diverse training and to emphasize or de-emphasize particular training data, it makes the optimization problem more challenging. In addition to the seven SNAP hyper-parameters in this three element system, the training set was sorted into groups of ground state, USPEX, surface and interfaces, room temperature AIMD, high temperature AIMD, m-active, and liquid structures. Furthermore, separate weights for energies and forces were provided for each training group. Given the large parameter optimization space, choosing the correct set of GA parameters in Dakota is key to expediting the optimization loop.  During the GA, we want to efficiently find SNAP candidates at the global minimum of error values rather than get trapped at a local minimum. We have found two GA parameters: population size and replacement type, have the greatest effect on the quality of output SNAP candidates. 
\par 
Large populations sizes require more computational time to reach subsequent generations, but allow more diversity in the population. Meanwhile, small population sizes evolve through generations more quickly but can lead to a poor solution \cite{Katoch2021}. To better determine ideal population size, we compared the quality of candidates from four, otherwise identical, Dakota runs with varying population size. Figure \ref{fig:pop_size} shows the total number of good candidates produced with respect to the total number of candidates evaluated over the course of each Dakota run. Good candidates were identified as those yielding errors below our desired accuracy for energies, forces, and objective functions. Looking at one optimization chronologically in Figure \ref{fig:pop_size}, a population size 100 (data in black circles), after four generations the number of these good candidates saturates at just 3, with no further improvement in the subsequent 60 generations. Increasing the population size to 500, 22 good candidates are found, though this run similarly, and unfortunately, saturated toward in the latter half of optimization. Tipping the balance ore toward exploration than exploitation, population size 1000 significantly improves the good candidate output to 81, and only begins to plateau toward the end of this run. Population size 1500 yields the greatest number of good candidates at 102, and we continue to see good candidates produced as the Dakota run progresses (i.e. no plateauing/converging on a local minima). However, there are diminishing returs of good candidates as the population size is increased to 2000. Here we can see how increasing the population size begins to slow down the progression through generations, resulting in fewer good candidates in the same amount of walltime. All subsequent optimizations were run with a population size of 1500.

\begin{figure}[htb]
    \centering
    \includegraphics[width=0.45\textwidth]{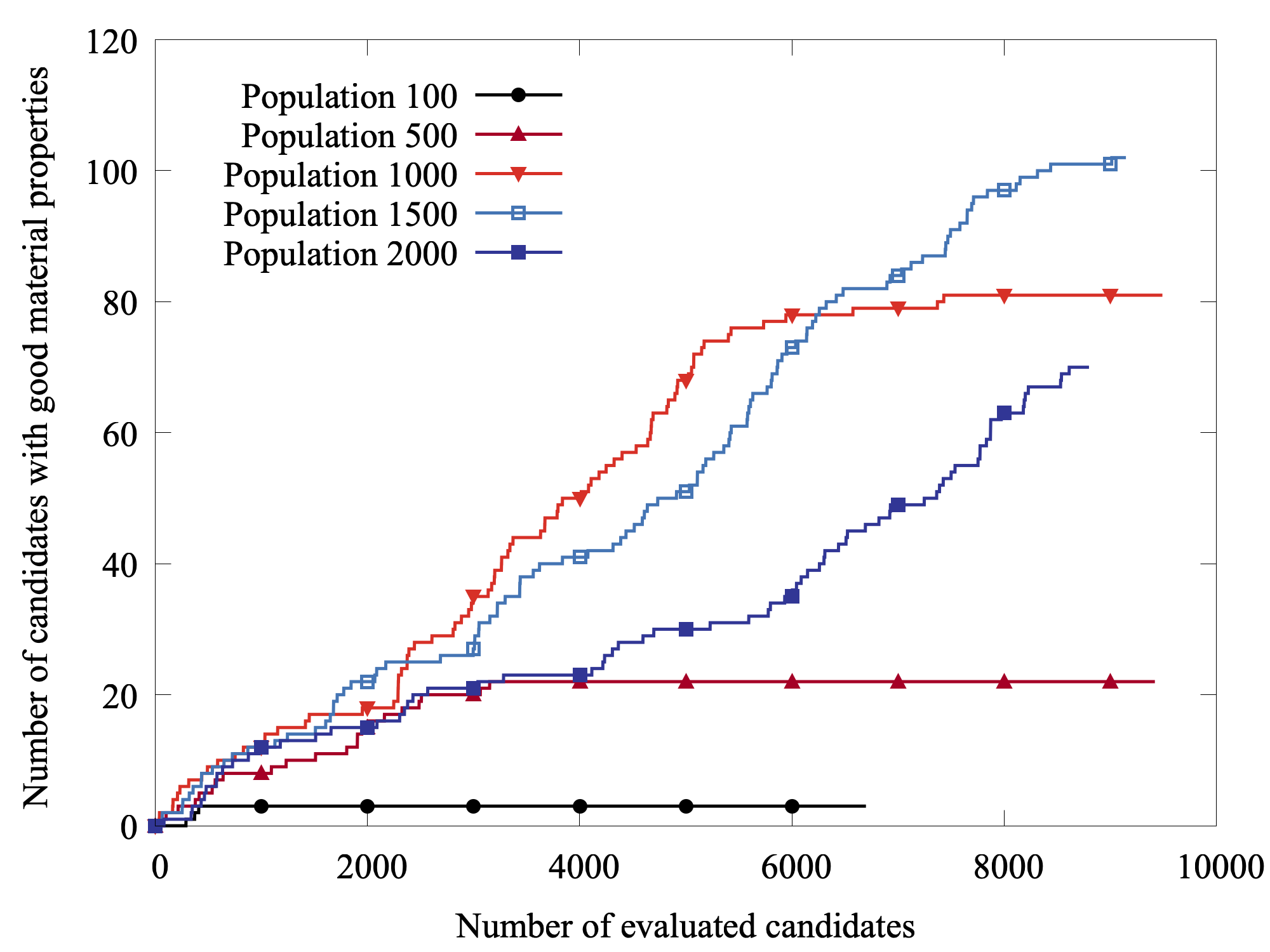}
    \caption{Number of good candidates produced from Dakota runs with respect to total number of evaluated candidates. The population size is varied from 100 to 2000. These runs work to optimize 21 variables, and all GA parameters aside from population size are consistent across runs. }
    \label{fig:pop_size}
\end{figure}

\par 
Over the course of the GA, loss of diversity in the candidate hyper-parameter values can also coincide with premature convergence \cite{Katoch2021}, or a result at a local rather than global minimum. This can be combatted by reducing the selection pressure which favors candidates with better fitness (lower errors) to act as parents for subsequent generations. To reduce the selection pressure, we used the ``favor feasible" replacement type in Dakota to allow any candidates with feasible errors to participate in reproduction. In contrast, the default replacement type ``elitist" only allows the best candidates to reproduce. ``Favor feasible" better maintains diversity in the candidate population, as shown in the t-SNE plot in Figure \ref{fig:favor_feasible_hyper_params}. This allows us to visualize how good candidates produced using different replacement types span hyper-parameter and group weight space. Figure \ref{fig:favor_feasible_hyper_params} shows that the candidates produced using the elitist method are limited to a smaller volume of parameter space. Meanwhile, the candidates produced using the favor feasible method span a larger volume and thus indicate greater diversity in the candidate parameters. Furthermore, from these otherwise identical runs, the elitist method produced 9 good candidates while the favor feasible method produced 27 good candidates.

\begin{figure}[htb]
    \centering
    \includegraphics[width=0.45\textwidth]{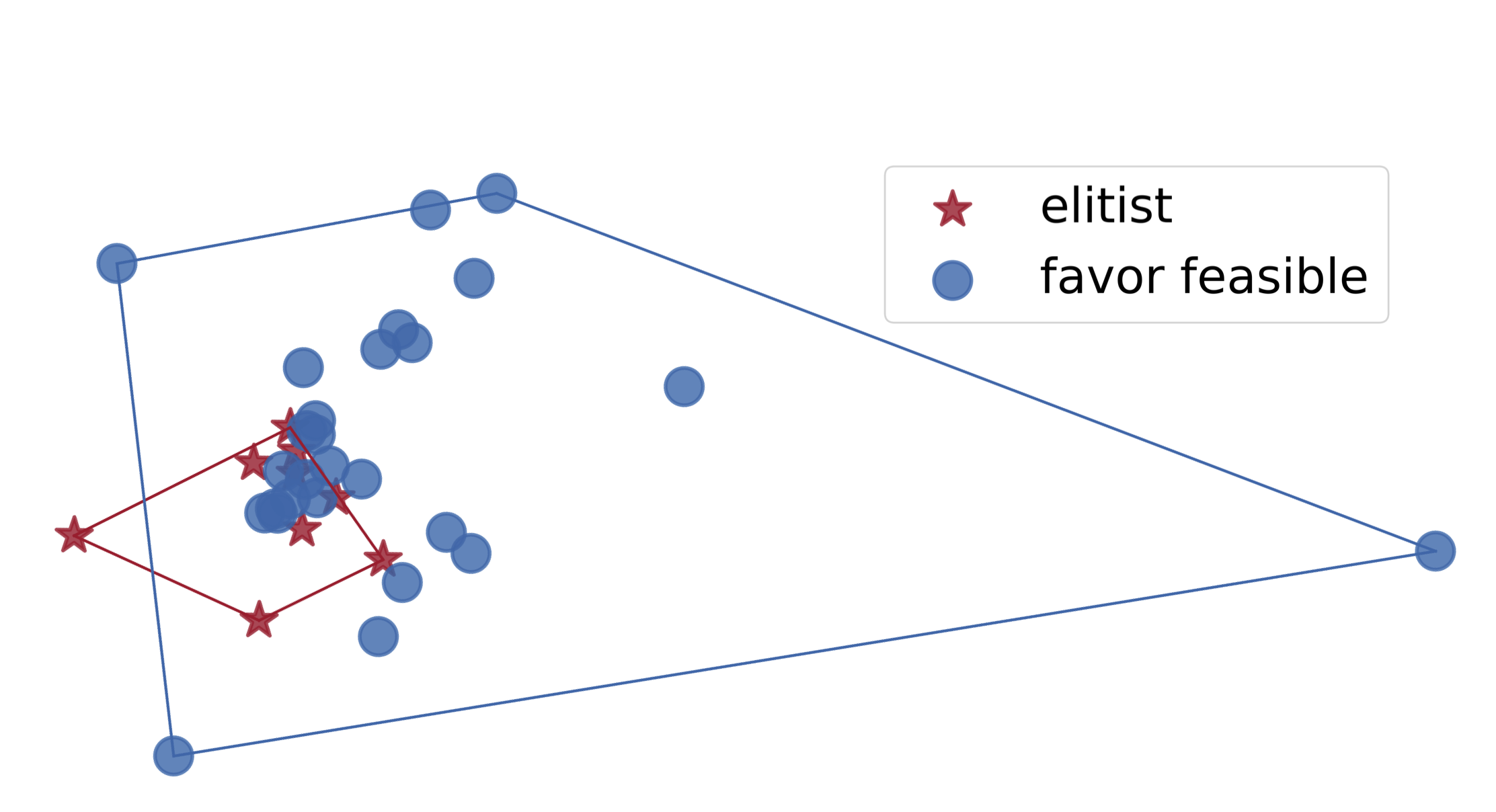}
    \caption{t-SNE of the 21 total hyper-parameters and group weights of good candidates from two Dakota runs. The red triangles and blue circles show candidates produced using the ``elitist" and ``favor feasible" replacement types, respectively. All other GA parameters, including a population size of 1500, were identical between the two runs.}
    \label{fig:favor_feasible_hyper_params}
\end{figure}

\par 

\subsubsection{\label{optimized-snap-vars}Optimized SNAP variables}

Using these objective functions and optimized GA parameters resulted in a production W-ZrC potential that can run at divertor temperatures. The final SNAP variables are shown in Table \ref{table:hyper-parameters}. Here we can see how Dakota prioritized different training groups with weights ranging from approximately 3 to 8. We note the high weight on ground state energies and corresponding low error. The surfaces and interfaces force weight also needs to be relatively high to account for the frozen layers in those simulations. Without a high force weight, SNAP may see frozen layers as stable and attempt to produce them during molecular dynamics simulations. 
\par






\renewcommand{\arraystretch}{1.5}
\setlength{\tabcolsep}{31pt}

\begin{table*}
\caption{Optimized SNAP hyper-parameters: element weights $w$, radial cutoffs $R$, and inner cutoff parameters $S$ and $D$. For each training group, the optimized energy weight (EW), optimized force weight (FW), energy mean absolute error (EMAE) (eV/atom), and force mean absolute error (FMAE) (eV/\AA) are given.}
\begin{tabular}{l c c c c}

\hline
\hline
& W & Zr & C &\\
\hline 
$w$ (unitless) & 1.0 & 0.63 & 1.05 & \\
$R$ (\AA) & 4.42 & 5.26 & 3.29 &\\
$S$ (\AA) & 0.88 & 0.88 & 0.88 &\\
$D$ (\AA) & 0.12 & 0.12 & 0.12 &\\
\hline
training group & EW & FW & EMAE & FMAE \\ 
\hline
ground state & 3.94 & 7.32 & 4.3 $\times$ 10$^{-2}$ & 3.8 $\times$ 10$^{-2}$\\

USPEX & 6.41 & 3.94 & 1.1 $\times$ 10$^{0}$ & 2.6 $\times$ 10$^{0}$\\

Surfaces and Interfaces & 3.36 & 6.52 & 8.9 $\times$ 10$^{-2}$ & 3.0 $\times$ 10$^{-1}$ \\

AIMD 300~K & 3.5 & 3.74 & 1.2 $\times$ 10$^{-1}$ & 5.3 $\times$ 10$^{-1}$ \\

AIMD $\geq$1000~K & 4.1 & 7.94 & 4.2 $\times$ 10$^{-1}$ & 1.0 $\times$ 10$^{-1}$\\

M-active & 3.3 & 3.62 & 8.8 $\times$ 10$^{-1}$ & 3.5 $\times$ 10$^{0}$ \\

Liquid & 3.68 & 5.36 & 2.7 $\times$ 10$^{-1}$  & 2.4 $\times$ 10$^{0}$\\

\hline
\hline

\label{table:hyper-parameters}
\end{tabular}

\end{table*}
\end{slow}

\subsection{\label{interpolation}Properties and Performance}

\renewcommand{\arraystretch}{1.5}
\setlength{\tabcolsep}{7pt}

 The predictions of our SNAP potential for several material properties are compared to DFT or experiment in Table \ref{table:dft-vs-snap}. These values are in good agreement, with surface stability ordering for the (100) and (110) surfaces in the appropriate order for both W and ZrC. The linear thermal expansion values for W and ZrC are given at the nearest experimentally available temperature to the maximum expected normal operation divertor temperature, taken from Miiller \textit{et. al.}\cite{Miiller1990} and Kostanovskiy \textit{et. al.}\cite{Kostanovskiy2018}, respectively.


\begin{table*}[]
\caption{Lattice parameter $a$ (\AA), bulk modulus $B$ (GPa), surface energies $E_{surf}$ (eV/\AA$^{2}$), and thermal expansion $\alpha$ predicted by SNAP as compared to reference values.}

\begin{tabular}{ l l l l l l l l l l l }
\hline 
\hline
& $a_{W}$ & $a_{ZrC}$ & $B_{W}$ & $B_{ZrC}$ & $E^{W(100)}_{surf}$ & $E^{W(110)}_{surf}$ & $E^{ZrC(100)}_{surf}$ & $E^{ZrC(110)}_{surf}$ & $\alpha^{2600~K}_{W}$ & $\alpha^{2573~K}_{ZrC}$\\
\hline
reference& 3.18$^{*}$ & 4.70$^{*}$ & 301.4$^{*}$ & 216.0$^{*}$ & 4.13$^{*}$ & 3.18$^{*}$ & 1.63$^{*}$ & 3.31$^{*}$ & 1.31$^{\dagger}$ & 1.85$^{\dagger}$ \\
SNAP & 3.19 & 4.78 & 303.3 & 209.0 & 3.38 & 3.22 & 1.40 & 2.75 & 1.05 & 1.50 \\
\hline

$^{*}$DFT\\
$^{\dagger}$experiment\cite{Miiller1990,Kostanovskiy2018}\\
\end{tabular}
\label{table:dft-vs-snap}
\end{table*}

\par 
To demonstrate the performance of the potential in predicting thermomechanical properties, we performed tensile tests on three W-ZrC interfaces: W(100)-ZrC(100), W(110)-ZrC(111) C-terminated, and W(110)-ZrC(111) Zr-terminated. These tests allow for comparison to interface predictions from DFT and extension to properties only attainable at larger length and time scales. Previous DFT studies have disagreed on which W-ZrC interface is the most stable. Zhang \textit{et. al.} \cite{Zhang2020} found W(100)-ZrC(100) to be the most stable, attributing the adhesion to strong W-C bonds at the interface. In contrast, Mukhopadhyay \textit{et. al.} reported that with more rigorous optimization the most stable interface is W(110)-ZrC(111) C-terminated. Despite the disagreement on most stable interface, both studies attribute interface stability to the strength of the covalent and ionic W-C bonds. By conducting tensile tests, we can compare ultimate tensile strength (UTS) to these DFT predictions.
\par 
The tensile test simulations were performed using the LAMMPS molecular dynamics package\cite{Plimpton1995,Thompson2022} and the SNAP potential developed in this work. The structure visualizations and atomic profiles were obtained using the Open Visualization Tool (OVITO)\cite{ovito}.
\par
\begin{figure}[htb]
    \centering
    \includegraphics[width=0.45\textwidth]{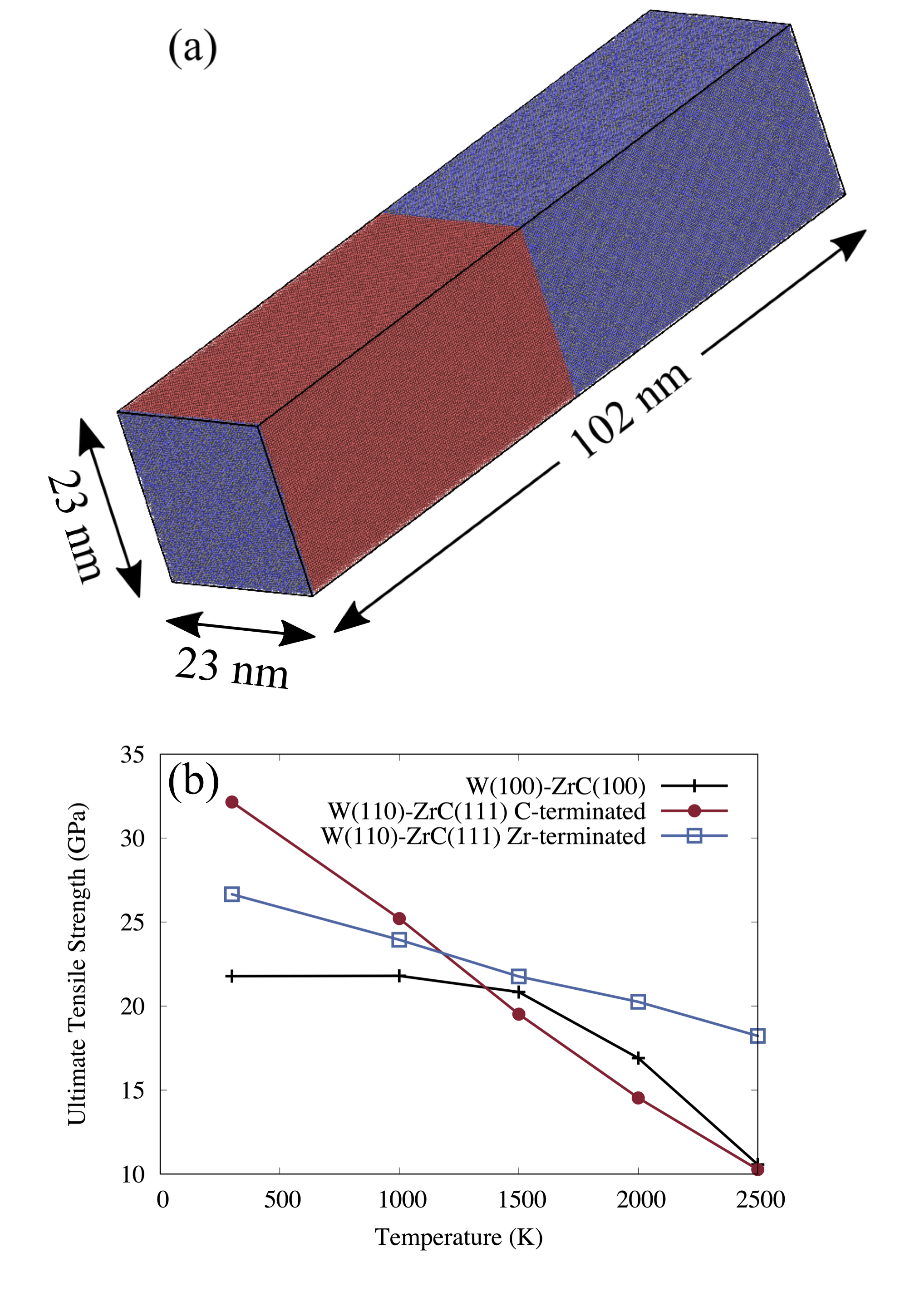}
    \caption{(a) Full view of the W(100)-ZrC(100) bicrystal after equilibration at 2500~K and (b) ultimate tensile strength with respect to temperature for each of the studied interfaces. W, Zr, and C atoms are colored red, blue, and grey, respectively.}
    \label{fig:tensile_strength}
\end{figure}

 
The bicrystal simulation cells were constructed to be approximately 20 nm $\times$ 100 nm $\times$ 20 nm, as shown in Figure \ref{fig:tensile_strength}(a). This corresponds to approximately 3.68 and 2.85 million atoms for the W(100)-ZrC(100) and W(110)-ZrC(111) interfaces, respectively. The bicrystals were first equilibrated at 50~K using an isobaric (constant number of atoms, pressure and temperature; NPT) ensemble. To produce equilibrated structures at each temperature of interest, the 50~K structures were heated to 100~K and then up to 2500~K in intervals of 100~K. At each temperature, dynamics were run with an NPT ensemble for 25 ps to allow for thermal expansion, followed by evolution in the canonical (constant number of atoms, volume and temperature; NVT) ensemble for 5 ps. For the tensile tests, the equilibrated structures were deformed in the interface-normal direction at a strain rate of 1$\times$10$^{8}$s$^{-1}$. 
\par 

\par 
The UTS from the tensile tests are shown in Figure \ref{fig:tensile_strength}(b). The full stress-strain curves are shown in Figure \ref{fig:stress-strain-curves}.
Failure of all interface types was observed through interface delamination, with limited plasticity observed prior, and as such we believe a direct comparison to DFT predicted interface cohesion is warranted.
At room temperature, our results agree with Mukhopadhyay \textit{et. al.} that the W(110)-ZrC(111) C-terminated interface has the strongest adhesion and thus highest UTS.
Conversely, the W(100)-ZrC(100) has the weakest adhesion, and the W(110)-ZrC(111) Zr-terminated interface lies between these limits. 
However, this ordering does not hold as temperature increases, with the W(110)-ZrC(111) C-terminated interface dropping to the lowest UTS at 2500~K. 
Our predictions show the dramatic decrease in interface strength for carbon terminated systems.
In contrast, the W(110)-ZrC(111) Zr-terminated interface exhibits the highest UTS at high temperature.

\par
\begin{figure}[htb]
    \centering
    \includegraphics[width=0.45\textwidth]{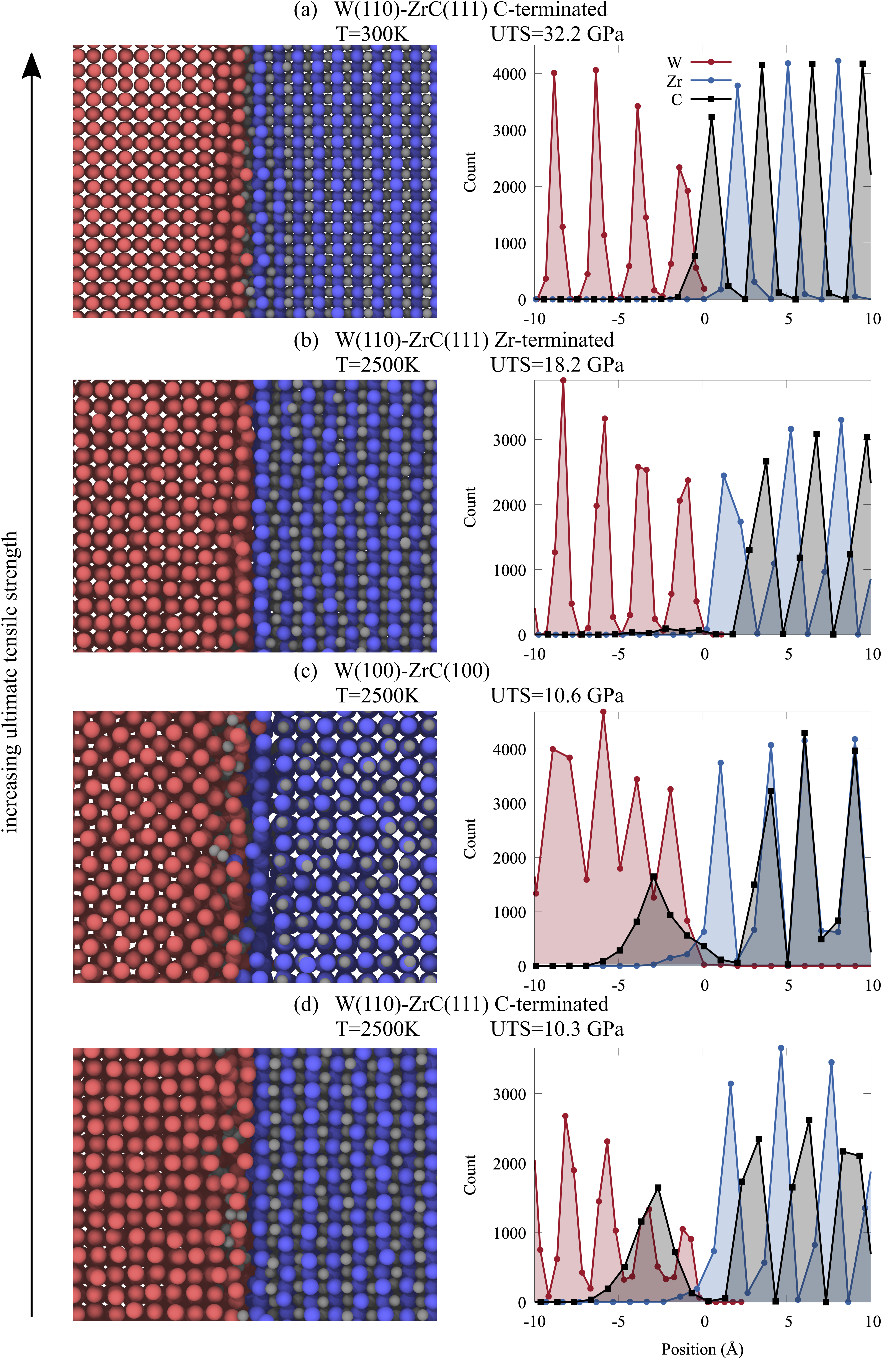}
    \caption{Atomic structures (left) and atomic profiles (right) of the investigated bicrystals at the W-ZrC interface immediately before fracture. The atomic structures are zoomed into a small portion of the interface to better visualize a sample of the composition and structure. Meanwhile, the atomic profiles show the total atom counts at the respective interface. The interface in the atomic profiles is shifted to $\approx$0 \AA. Atom colors are the same as those in Figure \ref{fig:tensile_strength}}
    \label{fig:atomic_profiles}
\end{figure}

To investigate atomistic mechanisms contributing to these differences in UTS with respect to interface and temperature, we examined the atomic structure and profiles across the interfaces just before failure, as shown in Figure \ref{fig:atomic_profiles}. For the strongest interface (Figure \ref{fig:atomic_profiles}(a)), the W(110)-ZrC(111) C-terminated at 300~K has a W-C interface, with a limited amount of C diffused into the terminating W layer. This is in agreement with DFT predictions that interface stability derives from strong W-C bonds. We also see the reconstruction of the C against the terminating layer of W, in agreement with Mukhopadhyay \textit{et. al.}\cite{Mukhopadhyay2022}. However, the atomic profile of the W(110)-ZrC(111) C-terminated interface does not remain constant as the bicrystal ramps. At 2500~K, the entire terminating C layer has diffused into the first three layers of the W (Figure \ref{fig:atomic_profiles}(d)). Even an amount of the now terminating Zr layer diffused into the W. 
This results in a W-Zr interface that exhibits significantly weaker UTS. 
The W(100)-ZrC(100) interface exhibits a very similar behavior at 2500~K (Figure \ref{fig:atomic_profiles}(c)), with nearly all first layer C and even some second layer C diffusing into the W.  
The W(110)-ZrC(111) Zr-terminated interface exhibits an intermediate behavior, in that it is weaker than the W(110)-ZrC(111) C-terminated 300~K interface, but it has the highest UTS of the studied interfaces at 2500~K. 
At the W(110)-ZrC(111) Zr-terminated interface at 2500~K, the interface is still W against Zr, with dilute amounts of C diffused into the W.
These diffuse interfaces are attributed to the overall decreasing strength at high temperatures, and should be a focus of experimental characterization of these proposed divertor materials. 
\par

\section{\label{discussion}Discussion}


As demonstrated herein MLIAPs offer first-principles accurate predictions at more experimentally relevant length and time scales. 
While best practices for developing MLIAPs are still evolving, the detailed procedure for training set curation and optimization captured here should be used as a guide for future studies. 
Our W-ZrC SNAP potential yields material properties in good agreement with reference values (Table \ref{table:dft-vs-snap}) while also delivering stable dynamics at fusion reactor temperatures. 
This was achieved in part by including DE structures calculated at traditional DFT rigor as well as high throughput, high temperature AIMD. 
Additional high energy, BDE structures were included to expand the application space where the potential can be used accurately. 
In this training set, the $>$200 atom high temperature AIMD and USPEX configurations most notably described regions of descriptor space BDE (Figure \ref{fig:t-sne}). 
Capturing complex chemical environments was achieved via USPEX variable composition configurations, which aided the potential in predicting diffusion, especially since such behavior was not prominent in the DE structures. 
Including this wide range of training data was enabled by the group weight flexibility in FitSNAP in conjunction with objective functions that held the potential to a high standard of both accuracy and stability. 
However, the group weight feature increases the parameter space that needs to be optimized. 
This search can be expedited by using efficient GA parameters as demonstrated here.  

Unsurprisingly, it is challenging to reach good agreement with all DFT energies, forces, material properties, and stability in a single SNAP potential. 
Furthermore, some SNAP candidates yielded very low energy, force, and material property errors but had unacceptable dynamics at expected divertor temperatures. This is in agreement with recent work that low force errors alone do not ensure an effective MLIAP\cite{Fu2022}. 
The production SNAP potential was selected as one capable of both reasonable errors and stable dynamics. 
We highly recommend using high temperature and/or stability objective functions or other means of ensuring stable dynamics for future MLIAPs with or without high temperature applications\cite{Novoselov2019}.
\par 
Our $\approx$3 million atom tensile tests support the findings of Mukhopadhyay et al. that with appropriate interface optimization DFT can predict the strongest interface. 
However, our study suggests that for the W-ZrC system this prediction is primarily valid only at room temperature. 
As the temperature increases, the atomic profile evolves as more terminating C diffuses into the W. 
This diffusion can only be observed at MD scales, and this is now possible to investigate further using the potential developed in this work. 
Our results motivate that future DFT studies may need to consider temperature effects when evaluating interfaces with high temperature applications.  
It is known the divertor region will experience a range of temperatures during regular operation which will affect the stability of the W-ZrC interfaces and therefore the strength of the divertor material. 
During normal operations, the W(110)-ZrC(111) C-terminated interface will have higher strength but this will likely deteriorate during an off-normal event like an ELM where the divertor will reach higher temperatures.  
In addition, the observed carbon diffusion from the interface into the tungsten matrix will change the initial interface structure, implying that at high temperatures, the interface will continue to evolve and potentially behave differently than the nominal material. 
These results indicate that the interface between the tungsten and dispersoid will need to be considered as a design choice and be optimized not just for normal operating conditions but for potential high temperature events at the divertor if the beneficial strength properties of these types of materials are to be properly quantified and maintained in a realistic operating environment.
Beyond the mechanical tests demonstrated here, W-ZrC dispersion strengthened materials will have a complex interplay with hydrogen retention.
Further MLIAP developments are needed to address this issue, but is well posed for classical MD as the time and length scales of retention/diffusion are congruent with the method. 

\section{\label{conclusions}Conclusions}

Developing a SNAP potential for fusion material requires careful curation of training data, selection of objective functions, and optimization of variables. DE and BDE structures were included in the training set for accurate reproduction of desired material properties and expanded accurate application space. Of note are the additional regions of descriptor space that can be covered with structures generated from high temperature $>$200 atom AIMD simulations and from USPEX variable composition searches. The production potential was reached using a variety of material property and stability objective functions as well as optimized genetic algorithm searches. Ultimately, the production W-ZrC SNAP potential provides a balance of good agreement with lattice parameters, surface energies, bulk moduli, thermal expansion, and stable dynamics at divertor temperatures. We demonstrated that the potential can be used to conduct millions of atoms simulations to test tensile strength, allowing for investigation of thermomechanical properties and atomic structure from 300~K - 2500~K. We found that while a W-C interface has the highest UTS, most C in the terminating layer of ZrC will diffuse into W at 2500~K. This results in a W-Zr interface with much lower UTS, implying the importance of quantifying the response and potential limitations of these materials over all fusion device operating temperatures. These tests were made possible by with an efficient, accurate MD potential. Now, the potential can be used to investigate dispersion-strengthened divertor microstructures at fusion reactor temperatures. 
This work has enabled the first IAP, of any kind, for the WZrC ternary and signals a promising future for the applicability of SNAP ML-IAP for complex materials.  
\par


\begin{acknowledgments}
We would like to thank Saikat Mukhopadhyay and Brian Wirth for sharing their atomic structures and helpful discussion relevant to this work. We also appreciate helpful feedback from James Goff for revising this manuscript. This work was supported by the U.S. Department of Energy, Office of Fusion Energy Sciences (OFES) under Field Work Proposal Number 20-023149.
This research used resources of the Oak Ridge Leadership Computing Facility, which is a DOE Office of Science User Facility supported under Contract DE-AC05-00OR22725. This article has been authored by an employee of National Technology \& Engineering Solutions of Sandia, LLC under Contract No. DE-NA0003525 with the U.S. Department of Energy (DOE). The employee owns all right, title and interest in and to the article and is solely responsible for its contents. The United States Government retains and the publisher, by accepting the article for publication, acknowledges that the United States Government retains a non-exclusive, paid-up, irrevocable, world-wide license to publish or reproduce the published form of this article or allow others to do so, for United States Government purposes. The DOE will provide public access to these results of federally sponsored research in accordance with the DOE Public Access Plan https://www.energy.gov/downloads/doe-public-access-plan.
\end{acknowledgments}

\section*{Author Declaration}
\subsection*{Conflict of Interest}
The authors have no conflicts to disclose.
\subsection*{Author Contributions}
E.L.S. and J.T. performed the DFT calculations. J.T. conceptualized the methodology and E.L.S. developed the production potential. M.A.C. and M.A.W. provided direction for the potential development and LAMMPS simulations. M.J.M automated the bicrystal generation and E.L.S. performed the MD simulations. E.L.S. wrote the original draft and all authors contributed to, reviewed, and edited the manuscript. A.P.T. and M.J.M. implemented the inner cutoff.  A.P.T. supervised the project.
\subsection*{Data Availability}
The data to produce the W-ZrC potential will be available at https://github.com/FitSNAP/FitSNAP upon publication.

\section*{\label{references}References}
\bibliography{WZrC.bib}
\clearpage


\renewcommand{\thepage}{S\arabic{page}} 
\renewcommand{\thesection}{S\arabic{section}}  
\renewcommand{\thetable}{S\arabic{table}}  
\renewcommand{\thefigure}{S\arabic{figure}} 
\setcounter{figure}{0}
\setcounter{section}{0}
\setcounter{page}{1}


\section*{Supporting Information}

\begin{figure}[h]
    \includegraphics[width=0.45\textwidth]{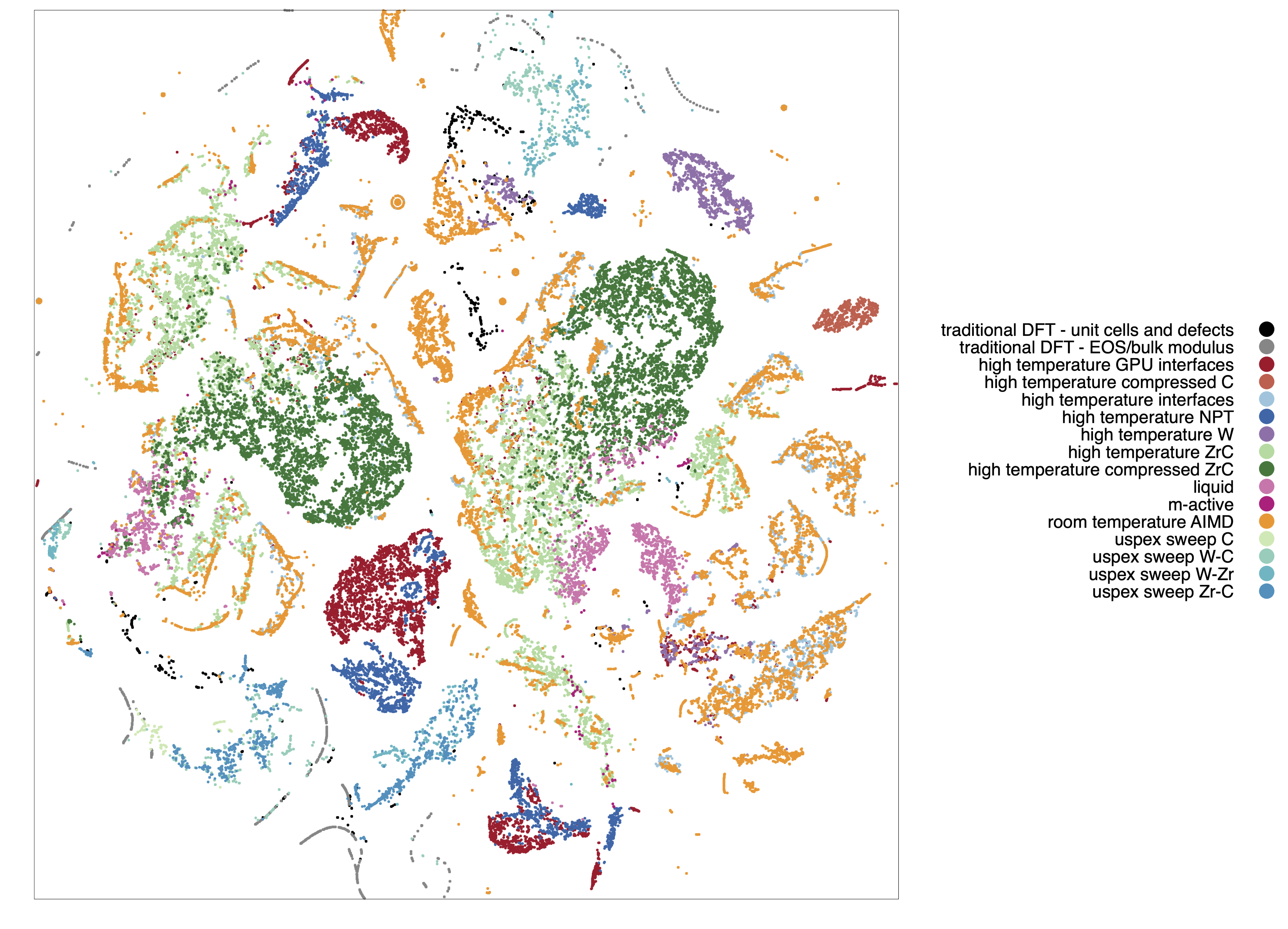}
    \caption{Visualization of training set coverage on the bispectrum descriptor space in 2D using distance preserving t-SNE analysis including all labels.}
    \label{fig:tsne-all-labels}
\end{figure}

\begin{figure}[h]
    \includegraphics[width=0.45\textwidth]{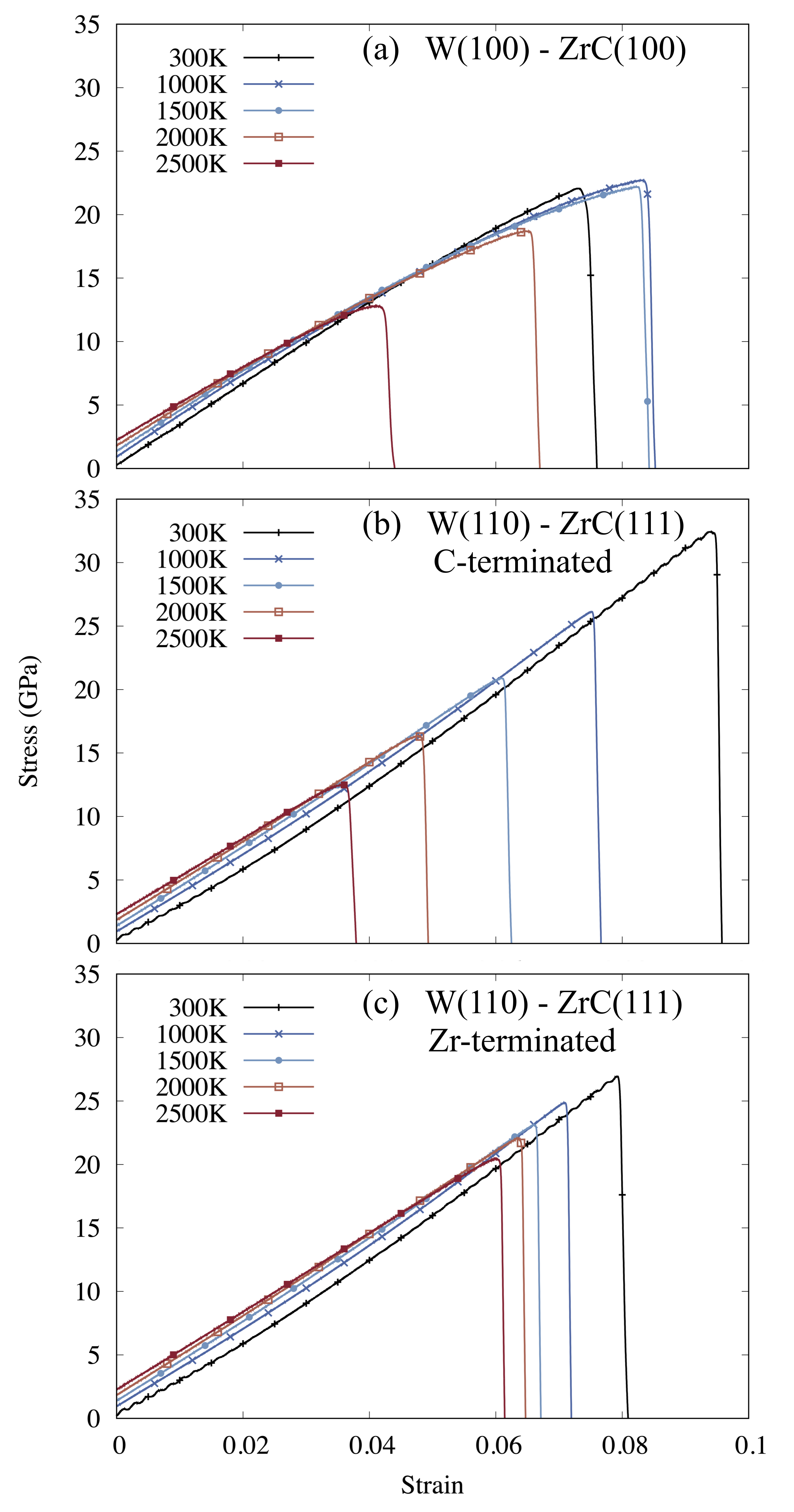}
    \caption{Stress strain curves for each of the studied interfaces.}
    \label{fig:stress-strain-curves}
\end{figure}

\end{document}